\documentclass[conference]{IEEEtran}
\def\BibTeX{{\rm B\kern-.05em{\sc i\kern-.025em b}\kern-.08em
    T\kern-.1667em\lower.7ex\hbox{E}\kern-.125emX}}
\usepackage{amsmath}
\usepackage{amsthm}
\usepackage{amsfonts}
\usepackage{amssymb}
\usepackage{graphicx}
\usepackage{cite}
\usepackage{subfigure}
\usepackage{balance}
\usepackage{algorithm}
\usepackage{algorithmic}
\usepackage{enumerate}
\usepackage{color}
\usepackage{array}
\usepackage{indentfirst}
\usepackage{multirow}
\usepackage{booktabs}
\usepackage{color}
\usepackage{slashbox}
\usepackage{diagbox}
\usepackage{threeparttable}
\usepackage[svgnames,table]{xcolor}
\usepackage{bm}
\usepackage{bbm}
\usepackage{subfigure}
\usepackage{epstopdf}
\usepackage{fancyhdr}

\makeatletter

\newcommand{\Rmnum}[1]{\expandafter\@slowromancap\romannumeral #1@}
\makeatother

\begin{document}
\title{ Deep Learning-Empowered Predictive Beamforming for IRS-Assisted Multi-User Communications
}



\author{Chang Liu$^{\ast}$, Xuemeng Liu$^{\S}$, Zhiqiang Wei$^{\ast}$, Shaokang Hu$^{\ast}$, Derrick Wing Kwan Ng$^{\ast}$, and Jinhong Yuan$^{\ast}$ \\
$^{\ast}$School of Electrical Engineering and Telecommunications, University of New South Wales, Sydney, Australia \\
$^{\S}$School of Electrical and Information Engineering, University of Sydney, Sydney, Australia \\
Email: $^{\ast}$\{chang.liu19, zhiqiang.wei, shaokang.hu, w.k.ng, j.yuan\}@unsw.edu.au, $^{\S}$xuemeng.liu@sydney.edu.au
}

\maketitle


\begin{abstract}
The realization of practical intelligent reflecting surface (IRS)-assisted multi-user communication (IRS-MUC) systems critically depends on the proper beamforming design exploiting accurate channel state information (CSI).
However, channel estimation (CE) in IRS-MUC systems requires a significantly large training overhead due to the numerous reflection elements involved in IRS.
In this paper, we adopt a deep learning approach to implicitly learn the historical channel features and directly predict the IRS phase shifts for the next time slot to maximize the average achievable  sum-rate of an IRS-MUC system taking into account the user mobility.
By doing this, only a low-dimension multiple-input single-output (MISO) CE is needed for transmit beamforming design, thus significantly reducing the CE overhead.
To this end, a location-aware convolutional long short-term memory network (LA-CLNet) is first developed to facilitate predictive beamforming at IRS, where the convolutional and recurrent units are jointly adopted to exploit both the spatial and temporal features of channels simultaneously.
Given the predictive IRS phase shift beamforming, an instantaneous CSI (ICSI)-aware fully-connected neural network (IA-FNN) is then proposed to optimize the transmit beamforming matrix at the access point.
Simulation results demonstrate that the sum-rate performance achieved by the proposed method approaches that of the genie-aided scheme with the full perfect ICSI.
\end{abstract}

\section{Introduction\label{sect: intr}}
Recently, intelligent reflecting surface (IRS) \cite{gong2020toward, zhang2020prospective} has been proposed as a promising technology for enabling future smart radio, due to its powerful capability in shaping wireless channels to improve the system data rate and communication reliability.
Generally, an IRS consists of a large number of passive and reconfigurable reflecting elements.
Through adapting the phase shifts of each reflecting element according to the channel conditions, an IRS can alter its reflection to the desired receivers.
Therefore, the design of IRS phase shifts configuration, i.e., passive beamforming, is essential for realizing practical IRS and has been widely investigated \cite{wu2021intelligent, hu2021robust, liu2020joint, wu2020towards}.

To fully unleash the potentials of IRS, numerous effective beamforming algorithms and schemes have been proposed for IRS-assisted communications \cite{li2021intelligent}.
For example, \cite{wu2019beamforming} proposed a joint active and passive beamforming scheme to minimize the total transmit power for IRS-enhanced communication networks.
Moreover, a block coordinate descent-based method was studied to maximize the weighted sum-rate of an IRS-aided system \cite{guo2020weighted}.
On the other hand, exploiting the powerful data-driven capability, the deep learning (DL)-based methods \cite{liu2020deeptransfer, lxm2020deepresidual, liu2020location, yuan2020learning, liu2019deep} have proven their effectiveness in IRS-assisted systems \cite{wang2021interplay}.
For instance, a deep deterministic policy gradient-based joint passive and active beamforming scheme was developed in \cite{huang2020reconfigurable} to maximize the sum-rate of IRS-assisted multiuser downlink multiple input single-output systems, which can achieve a comparable performance with the state-of-the-art schemes.
Also, in \cite{gao2020unsupervised}, an unsupervised DL approach was proposed for phase shift optimization and presented a decent rate performance requiring only a low complexity.
Note that the aforementioned beamforming studies, i.e., \cite{wu2019beamforming, guo2020weighted, wang2021interplay, huang2020reconfigurable, gao2020unsupervised}, optimistically assumed the availability of perfect channel estimation (CE) for beamforming designs.
However, since an IRS generally consists of numerous reflection elements, existing CE schemes either have limited CE accuracy or require exceedingly large training overheads \cite{liu2020deepresidual}, which remain a major obstacle for realizing IRS in practice.

To strike a balance between the required CE signaling overhead and the beamforming performance, in this paper, we propose an unsupervised DL-based predictive beamforming scheme to implicitly learn the historical channel features for establishing the IRS phase shifts to maximize the average sum-rate of an IRS-assisted multi-user communication (IRS-MUC) system.
In the proposed scheme, a location-aware convolutional long short-term memory network (LA-CLNet) is first developed to predict the IRS phase shifts to be adopted for the next time slot, where both the convolutional and recurrent units are jointly adopted to exploit the spatial and temporal features of the historical line-of-sight channels.
Based on the predictive IRS phase shifts, only an equivalent multiple-input single-output (MISO) CE is needed for obtaining the instantaneous CSI (ICSI). Then, an ICSI-aware fully-connected neural network (IA-FNN) is proposed to optimize the instantaneous beamforming matrix at the access point (AP).
Simulation results show that the sum-rate performance of the proposed method approaches that of the perfect genie-aided scheme exploiting the full ICSI.

\emph{Notations}:
Superscripts $T$ and $H$ represent the transpose and the conjugate transpose, respectively. $\mathbb{N}_1$, $\mathbb{C}$, and $\mathbb{R}$ are the sets of nonzero natural numbers, complex numbers, and real numbers, respectively. ${\mathcal{N}}( \bm{\mu},\mathbf{\Sigma} )$, and ${\mathcal{CN}}( \bm{\mu},\mathbf{\Sigma} )$ denote the Gaussian and circularly symmetric complex Gaussian distributions, where $\bm{\mu}$ and $\mathbf{\Sigma}$ are the mean vector and the covariance matrix, respectively.
$\arccos(\cdot)$ is the inverse cosine function.
$\mathcal{U}(a,b)$ denotes the uniform distribution within $[a,b]$.
$|\cdot|$ and $\|\cdot\|$ denote the absolute value and the Euclidian norm, respectively.
$E\{\cdot\}$ is the statistical expectation operation.
$\mathrm{diag}(\cdot)$ denotes generating a diagonal matrix and $\otimes$ is the Kronecker product.
$\mathrm{Re}\{\cdot\}$ and $\mathrm{Im}\{\cdot\}$ denote the real part and the imaginary part of a complex-valued input matrix, respectively.


\begin{figure}[t]
  \centering
  \includegraphics[width=\linewidth]{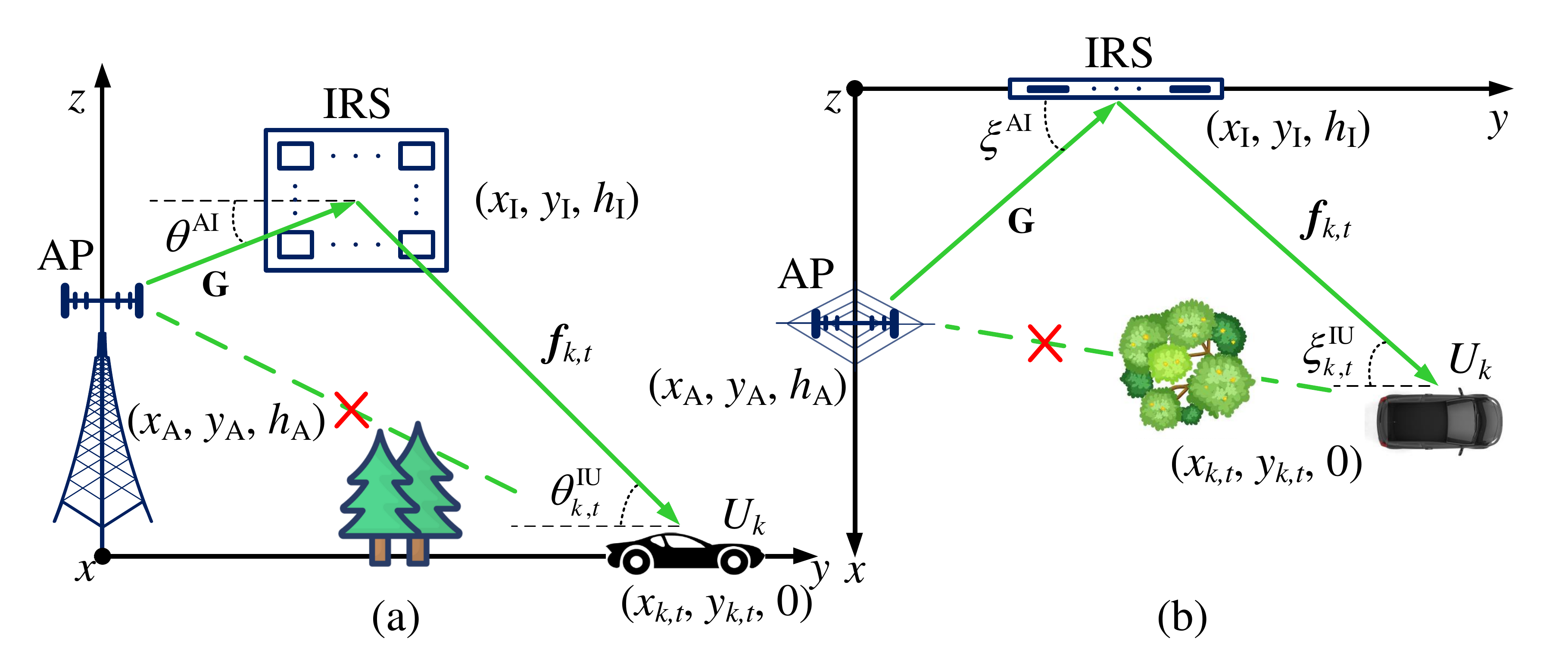}
  \caption{ The downlink of the considered IRS-MUC system with $K=1$ for illustration.
  (a) Side view; (b) Bird's-eye view.}\label{Fig:scenario_vertical_horizontal} 
\end{figure}

\section{System Model}
As shown in Fig. \ref{Fig:scenario_vertical_horizontal}, in this paper, we consider a downlink IRS-MUC system, where one AP equipped with $M$-element antenna array serves $K \geq 1$  single-antenna users with the assistance of an IRS with $N = N_y \times N_z$ passive reflecting elements.
Denote by $\mathcal{K} \triangleq \{1,2,\cdots,K\}$ the set of user index, then we adopt $U_k$, $k \in \mathcal{K}$, to represent the $k$-th user.
Specifically, we focus on a dynamic scenario where each user is in motion while the AP and the IRS are static.
In this case, the three-dimensional (3D) Cartesian coordinates of AP and IRS can be expressed as $\mathbf{L}_\mathrm{A} = [x_\mathrm{A},y_\mathrm{A},h_\mathrm{A}]^T$ and $\mathbf{L}_\mathrm{I} = [x_\mathrm{I},y_\mathrm{I},h_\mathrm{I}]^T$, respectively. Here, $x_\mathrm{A}$, $y_\mathrm{A}$, and $h_\mathrm{A}$ (or $x_\mathrm{I}$, $y_\mathrm{I}$,  and $h_\mathrm{I}$) are the coordinates of AP (or IRS) on $x$-axis, $y$-axis, and $z$-axis, respectively.
In addition, the location of $U_k$ at the $t$-th, $t \in \mathbb{N}_1$, time slot is $\mathbf{L}_{k,t} = [x_{k,t},y_{k,t},0]^T$, where $x_{k,t}$ and $y_{k,t}$ are the coordinates on $x$-axis, $y$-axis, respectively.
Thus, the distances of the AP-to-IRS and the IRS-to-$U_k$ at the $t$-th time slot can be expressed as
$d^\mathrm{AI} = \sqrt{(x_\mathrm{A} - x_\mathrm{I})^2 + (y_\mathrm{A} - y_\mathrm{I})^2 + (h_\mathrm{A} - h_\mathrm{I})^2}$ and $d^\mathrm{IU}_{k,t} = \sqrt{ (x_\mathrm{I} - x_{k,t})^2 + (y_\mathrm{I} - y_{k,t})^2 }$, respectively.
Specifically, the location of each user is determined by
\begin{equation}\label{movement_function}
  \mathbf{L}_{k,t+1} = \mathbf{L}_{k,t} + \mathbf{v}_{k,t} \Delta T + \mathbf{\Lambda}_{k,t}, \forall k,t.
\end{equation}
Here, $\mathbf{v}_{k,t} = [v_{k,t}^x, v_{k,t}^y, 0]^T$ is the average velocity of $U_k$ at the $t$-th time slot, where $v_{k,t}^x$ and $v_{k,t}^y$ are the velocity projections on the $x$-axis and $y$-axis, respectively.
Without loss of generality, we assume that both the amplitude $a = \|\mathbf{v}_{k,t}\|$ and the phase $b = \mathrm{arccos} (\frac{v_{k,t}^x}{\|\mathbf{v}_{k,t}\|})$ follow the uniform distributions, i.e., $a \sim \mathcal{U}(A_1,A_2)$ and $b \sim \mathcal{U}(B_1,B_2)$, where $A_1$, $A_2$, $B_1$, $B_2$ are constants with $A_1 \leq A_2$ and $B_1 \leq B_2$ \cite{9076668}.
In addition, $\Delta T$ is the duration of each time slot\footnotemark\footnotetext{We assume that $\Delta T$ is sufficiently short such that $\mathbf{L}_{k,t}$ is a constant. } and $\mathbf{\Lambda}_{k, t} = [\lambda_{k,t}^x, \lambda_{k,t}^y, \lambda_{k,t}^z]^T$ models the environment uncertainty, where $\lambda_{k,t}^\varepsilon \sim \mathcal{N}(0,\sigma_\lambda^2)$ with $\varepsilon = \{x,y,z\}$ denotes the uncertainty offset on the $\varepsilon$-axis and $\sigma_\lambda^2$ is the uncertainty variance.
For any time slot $t$, we assume that only the historical locations are known and this information will be adopted by the proposed DL approach for predictive beamforming.
Next, we will introduce the channel model, the signal model, and the transmission protocol, respectively.

\begin{figure*}[t]
  \centering
  \includegraphics[width=\linewidth]{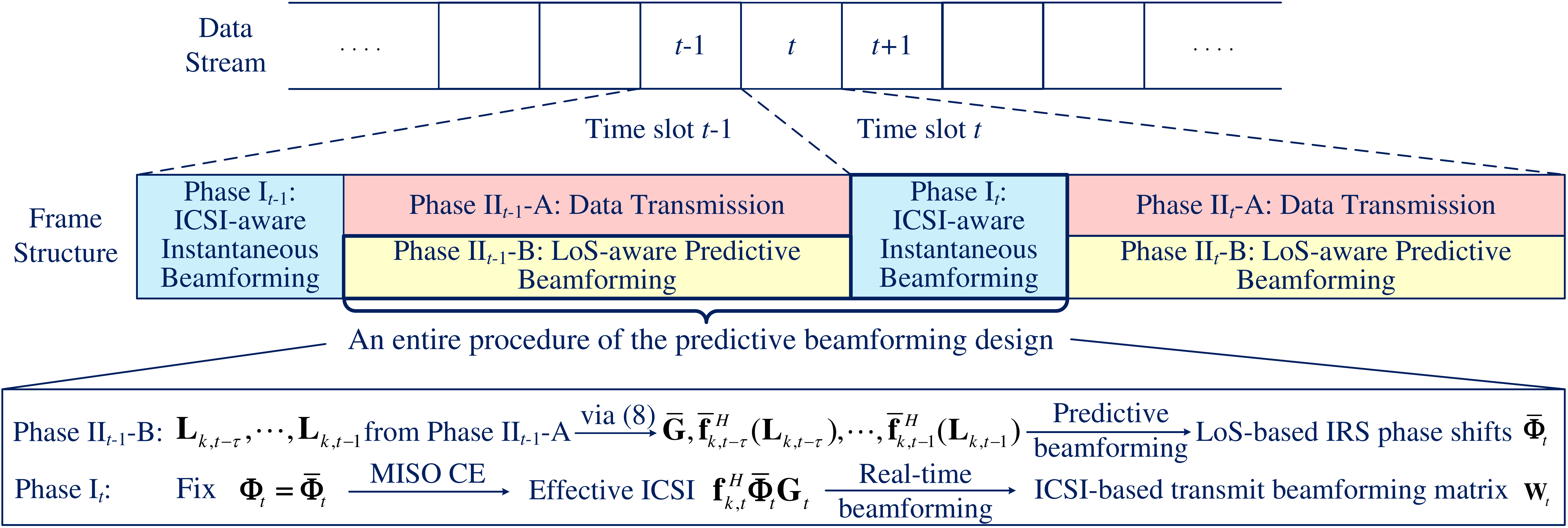}
  \caption{ The frame structure of the developed transmission protocol for (downlink) IRS-MUC systems.
  }\label{Fig:tranmission protocol}
\end{figure*}

\subsection{Channel Model}
As depicted in Fig. \ref{Fig:scenario_vertical_horizontal}, the direct AP-to-users links are assumed to be neglected due to the existence of unfavorable signal blockages, which is commonly adopted in e.g., \cite{huang2020reconfigurable, you2020energy}.
Since the LoS component and the non-LoS (NLoS) component may simultaneously exist in practical channels, we adopt a Rician fading model for all the channels.
In this case, the channel of the AP-to-IRS link can be formulated as \vspace{-0.05cm}
\begin{equation}\label{G_model}
  \mathbf{G}_t = \sqrt{\frac{\beta^{\mathrm{AI}}}{\beta^{\mathrm{AI}} + 1}}\bar{\mathbf{G}} + \sqrt{\frac{1}{\beta^{\mathrm{AI}} + 1}}\tilde{\mathbf{G}}_t.
  \vspace{-0.05cm}
\end{equation}
Here, $\beta^\mathrm{AI} \geq 0$ is the Rician factor of the AP-to-IRS link and $\tilde{\mathbf{G}}_t \in \mathbb{C}^{N \times M}$ denotes the NLoS component.
In addition, \vspace{-0.1cm}
\begin{equation}\label{G_bar}
  \bar{\mathbf{G}} = \mathbf{a}^\mathrm{AP}(\theta^\mathrm{AI}, \xi^\mathrm{AI}) \otimes \mathbf{a}^\mathrm{IRS}_y(\theta^\mathrm{AI}, \xi^\mathrm{AI}) \otimes \mathbf{a}^\mathrm{IRS}_z(\theta^\mathrm{AI}, \xi^\mathrm{AI})
\end{equation}
denotes the LoS component, where \vspace{-0.1cm}
\begin{equation}\label{}
\begin{aligned}
  \mathbf{a}^\mathrm{AP}(\theta^\mathrm{AI}, \xi^\mathrm{AI}) = [ &  1, e^{-j\frac{2\pi\Delta d_\mathrm{A}}{\lambda_c} \sin(\theta^\mathrm{AI})\cos(\xi^\mathrm{AI})}, \cdots,   \\
  & e^{-j\frac{2\pi (M -1) \Delta d_\mathrm{A}}{\lambda_c} \sin(\theta^\mathrm{AI})\cos(\xi^\mathrm{AI})}]^T ,
\end{aligned}
\end{equation}
\begin{equation}\label{}
\begin{aligned}
  \mathbf{a}^\mathrm{IRS}_y(\theta^\mathrm{AI}, \xi^\mathrm{AI}) = [ &  1, e^{-j\frac{2\pi\Delta d_{\mathrm{I}y}}{\lambda_c} \sin(\theta^\mathrm{AI})\cos(\xi^\mathrm{AI})}, \cdots,   \\
  & e^{-j\frac{2\pi (N_y -1) \Delta d_{\mathrm{I}y}}{\lambda_c} \sin(\theta^\mathrm{AI})\cos(\xi^\mathrm{AI})}]^T ,
\end{aligned}
\end{equation}
and \vspace{-0.2cm}
\begin{equation}\label{}
\begin{aligned}
  \mathbf{a}^\mathrm{IRS}_z(\theta^\mathrm{AI}, \xi^\mathrm{AI}) = [ &   1, e^{-j\frac{2\pi\Delta d_{\mathrm{I}z}}{\lambda_c} \sin(\theta^\mathrm{AI})\sin(\xi^\mathrm{AI})}, \cdots,   \\
 &  e^{-j\frac{2\pi (N_z -1) \Delta d_{\mathrm{I}z}}{\lambda_c} \sin(\theta^\mathrm{AI})\sin(\xi^\mathrm{AI})}]^T
 \vspace{-0.2cm}
\end{aligned}
\end{equation}
are the steering vectors of AP, IRS on $y$-axis, and IRS on $z$-axis, respectively.
Parameters $\theta^\mathrm{AI}$ and $\xi^\mathrm{AI}$ denote the horizontal and vertical angle-of-arrivals (AoAs) from the AP to the IRS, respectively.
As shown in Fig. \ref{Fig:scenario_vertical_horizontal}, we have $\sin(\theta^\mathrm{AI}) = \frac{|h_\mathrm{I} - h_\mathrm{A}|}{d^\mathrm{AI}}$, $\cos(\xi^\mathrm{AI}) = \frac{|y_\mathrm{I}|}{d^\mathrm{AI}}$, and $\sin(\xi^\mathrm{AI}) = \frac{|x_\mathrm{A}|}{d^\mathrm{AI}}$.
Besides, $\lambda_c$ denotes the wavelength of the carrier frequency, $\Delta d_\mathrm{A}$ is the distance between two adjacent antenna elements of AP, $\Delta d_{\mathrm{I}y}$ and $\Delta d_{\mathrm{I}z}$ are the distances between two adjacent IRS elements on $y$-axis and $z$-axis, respectively.
Similarly, the channel of the IRS-to-$U_k$ link at the $t$-th time slot can be expressed as
\begin{equation}\label{f_model}
  \mathbf{f}_{k,t} = \sqrt{\frac{\beta^{\mathrm{IU}}_{k,t}}{\beta^\mathrm{IU}_{k,t} + 1}}\bar{\mathbf{f}}_{k,t}(\mathbf{L}_{k,\lambda}) + \sqrt{\frac{1}{\beta^\mathrm{IU}_{k,t} + 1}}\tilde{\mathbf{f}}_{k,t}.
\end{equation}
Here, $\beta^\mathrm{IU}_{k,t} \geq 0$ is the Rician factor of the IRS-to-$U_k$ link and $\tilde{\mathbf{f}}_{k,t} \in \mathbb{C}^{N \times 1}$ denotes the NLoS component.
In addition,
\begin{equation}\label{f_bar}
  \bar{\mathbf{f}}_{k,t}(\mathbf{L}_{k,t}) = \mathbf{a}^\mathrm{IRS}_y(\theta^\mathrm{IU}, \xi^\mathrm{IU}) \otimes \mathbf{a}^\mathrm{IRS}_z(\theta^\mathrm{IU}, \xi^\mathrm{IU})
\end{equation}
is the LoS component as well as the function of $\mathbf{L}_{k,t}$, where
\begin{equation}\label{}
\begin{aligned}
  \mathbf{a}^\mathrm{IRS}_y(\theta^\mathrm{IU}_{k,t}, \xi^\mathrm{IU}_{k,t}) = [ &  1, e^{-j\frac{2\pi\Delta d_{\mathrm{I}y}}{\lambda_c} \sin(\theta^\mathrm{IU}_{k,t})\cos(\xi^\mathrm{IU}_{k,t})}, \cdots,   \\
  & e^{-j\frac{2\pi (N_y -1) \Delta d_{\mathrm{I}y}}{\lambda_c} \sin(\theta^\mathrm{IU}_{k,t})\cos(\xi^\mathrm{IU}_{k,t})}]^T
\end{aligned}
\end{equation}
\vspace{-0.1cm}
and
\begin{equation}\label{}
\begin{aligned}
  \mathbf{a}^\mathrm{IRS}_z(\theta^\mathrm{IU}_{k,t}, \xi^\mathrm{IU}_{k,t}) = [ &  1, e^{-j\frac{2\pi\Delta d_{\mathrm{I}z}}{\lambda_c} \sin(\theta^\mathrm{IU}_{k,t})\sin(\xi^\mathrm{IU}_{k,t})}, \cdots,   \\
  & e^{-j\frac{2\pi (N_z -1) \Delta d_{\mathrm{I}z}}{\lambda_c} \sin(\theta^\mathrm{IU}_{k,t})\sin(\xi^\mathrm{IU}_{k,t})}]^T
\end{aligned}
\end{equation}
are the steering vectors of IRS with
$\sin(\theta^\mathrm{IU}_{k,t}) = \frac{|h_\mathrm{I}|}{d^\mathrm{IU}_{k,t}}$, $\cos(\xi^\mathrm{IU}_{k,t}) = \frac{|y_{k,t} - y_\mathrm{I}|}{d^\mathrm{IU}_{k,t}}$, and $\sin(\xi^\mathrm{IU}_{k,t}) = \frac{|x_{k,t}|}{d^\mathrm{IU}_{k,t}}$, as illustrated in Fig. \ref{Fig:scenario_vertical_horizontal}.


\subsection{Signal Model}
Note that the IRS can reflect the incident signals to the desired users via controlling the phase shift matrix, i.e., $\mathbf{\Phi}_t = \mathrm{diag}([ e^{j\varphi_{1,t}}, e^{j\varphi_{2,t}},\cdots, e^{j\varphi_{N,t}} ]^T) \in \mathbb{C}^{N \times N}$, where $\varphi_{n,t}$ denotes the phase shift of the $n$-th, $n \in \mathcal{N} = \{1,2,\cdots,N\}$, IRS element at the $t$-th time slot.
In this case, the received signal in $U_k$ at the $t$-th time slot can be expressed as
\begin{equation}\label{G_bar}
  r_{k,t} = \mathbf{f}_{k,t}^H \mathbf{\Phi}_t \mathbf{G}_t\sum_{i=1}^{K}\mathbf{w}_{i,t}s_{i,t} + n_{k,t}, \forall t,k.
\end{equation}
Here, $\mathbf{w}_{i,t} \in \mathbb{C}^{M \times 1}$ and $s_{i,t} \sim \mathcal{CN}(0,1)$ denote the beamforming vector and the transmitted data for $U_i$ at the $t$-th time slot, respectively.
$n_{k,t} \sim \mathcal{CN}(0,\sigma_k^2)$ is the additive Gaussian noise sample in $U_k$ at the $t$-th time slot, where $\sigma_k^2$ is the noise variance at the receiver of $U_k$.
Based on this, the received signal-to-interference-plus-noise ratio (SINR) can be expressed as
\begin{align}\label{SINR}
  \gamma_{k,t}(\mathbf{\Phi}_t,\mathbf{w}_{k,t}) = \frac{|\mathbf{f}_{k,t}^H\mathbf{\Phi}_t\mathbf{G}_t\mathbf{w}_{k,t}|^2}{\sum_{j \neq k}^K|\mathbf{f}_{k,t}^H\mathbf{\Phi}_t\mathbf{G}_t\mathbf{w}_{j,t}|^2 + \sigma_k^2}.
\end{align}

\subsection{Transmission Protocol}
As mentioned in the introduction, the CE in IRS-MUC systems generally incurs a large training overhead since the IRS-associated cascaded channel, i.e., $\mathrm{diag}(\mathbf{f}_{k,t}^H) \mathbf{G}_t \in \mathbb{C}^{N \times M}$  is with a high dimension.
In fact, if the IRS phase shifts are preset, the required ICSI reduces to a low-dimensional MISO channel, i.e., $\mathbf{f}_{k,t}^H\mathbf{\Phi}_t\mathbf{G}_t \in \mathbb{C}^{1 \times M}$, thus the CE overhead is $1/N$ times lower than that of $\mathrm{diag}(\mathbf{f}_{k,t}^H) \mathbf{G}_t$.
Inspired by this, we propose a hierarchical transmission protocol, where the phase shift matrix at IRS adopted in next time slot has been properly predicted and fixed in advance such that only a MISO CE is required for optimizing the transmit beamforming.

As shown in Fig. \ref{Fig:tranmission protocol}, the developed transmission protocol consists of three phases: Phase I$_t$: ICSI-aware instantaneous beamforming, Phase II$_t$-A: Data transmission, and Phase II$_t$-B: LoS-aware predictive beamforming, where $t$ denotes the $t$-th time slot.
We assume that the channel coefficients and the locations of users change over time slots, while they remain static within each time slot.
Denote by $t-1$ the current time slot, an entire procedure of the predictive beamforming design for the $t$-th time slot consists of (i) Phase II$_{t-1}$-B and (ii) Phase I$_{t}$.
In (i), the AP obtains historical locations, i.e., $\mathbf{L}_{k,t-\tau}, \cdots, \mathbf{L}_{k,t-1}$, $\forall k$, through previous uplink transmission and assembles the historical LoS channels $ \{\bar{\mathbf{G}},\bar{\mathbf{f}}_{k,t-\tau}(\mathbf{L}_{k,t-\tau}), \cdots,\bar{\mathbf{f}}_{k,t-1}(\mathbf{L}_{k,t-1}) \}$, $\forall k$, via (\ref{f_bar}), where $\tau$ is the number of available historical time slots. Then, the AP predicts the LoS-based IRS phase shift matrix $\bar{\mathbf{\Phi}}_t$ to be adopted in time slot $t$.
In (ii), the IRS first sets the phase shift matrix as $\mathbf{\Phi}_t = \bar{\mathbf{\Phi}}_t$ for CE to acquire the effective ICSI $\mathbf{f}_{k,t}^H\bar{\mathbf{\Phi}}_t\mathbf{G}_t$. Then, this information is exploited to optimize the ICSI-based transmit beamforming matrix $\mathbf{W}_t$.
By doing this, the proposed protocol can reduce the CE overhead as well as maintain the beamforming performance.\vspace{-0.2cm}

\section{Problem Formulation}
In this paper, we aim to maximize the average sum-rate by jointly optimizing the LoS-aware phase shift matrix at IRS and the ICSI-aware beamforming matrix at AP subject to the power constraint at AP and the phase shift constraint at IRS.
Thus, for any time slot $t \in \mathcal{T} = \{t| t \geq \tau + 1, t \in \mathbb{N}_1 \}$, the optimization problem is formulated as \vspace{-0.2cm}
\begin{align}
\max_{\bar{\mathbf{\Phi}}_t} ~ &\mathbb{E}_{\bar{\mathbf{H}}_{t}|{\mathcal{H}}_t^{\tau}}\left\{\max_{{\mathbf{W}}_t}
\sum_{k = 1}^K\mathrm{log}_2\left(1 + \gamma_{k,t}(\bar{\mathbf{\Phi}}_t,\mathbf{w}_{k,t}) \right)\right\} \label{P1} \\
\mathrm{s.t.}\,\,&\sum_{k = 1}^K\|\mathbf{w}_{k,t}\|^2\leq P, ~ 0 \leq \bar{\varphi}_{n,t} \leq 2\pi, \forall n \in \mathcal{N}, t \in \mathcal{T}. \notag
\end{align}
Here, $P$ is the instantaneous maximum power budget at the AP and $\bar{\varphi}_{n,t}$ denotes the LoS-aware phase shift of the $n$-th element at the $t$-th time slot.
In addition, $\bar{\mathbf{\Phi}}_{t} = \mathrm{diag}([ e^{j\bar{\varphi}_{1,t}}, e^{j\bar{\varphi}_{2,t}},\cdots, e^{j\bar{\varphi}_{N,t}} ]) $ and $\mathbf{W}_t = [\mathbf{w}_{1,t},\mathbf{w}_{2,t},\cdots,\mathbf{w}_{K,t} ] $ denote the LoS-aware predictive phase shift matrix and the ICSI-aware beamforming matrix for time slot $t$, respectively.
Also, $\gamma_{k,t}(\cdot,\cdot)$ is the received SINR defined in (\ref{SINR}).
Note that (\ref{P1}) is a nested optimization problem. In particular, the inner maximization is over the ICSI-aware ${\mathbf{W}}_t$ with the optimized $\bar{\mathbf{\Phi}}_t$ from the outer maximization.
Meanwhile, the outer maximization is over the LoS-aware predictive $\bar{\mathbf{\Phi}}_{t}$ and the expectation $\mathbb{E}_{\bar{\mathbf{H}}_{t}|\mathcal{H}_t^{\tau}}\{\cdot\}$ is taken over all random realizations of $\bar{\mathbf{H}}_{t}$ given the historical LoS channels $\mathcal{H}_t^{\tau} \triangleq \{\bar{\mathbf{H}}_{t-1},\cdots,\bar{\mathbf{H}}_{t-\tau}\}$, where $\bar{\mathbf{H}}_{\lambda} \triangleq \{\bar{\mathbf{G}},\bar{\mathbf{f}}_{1,\lambda}(\mathbf{L}_{k,\lambda}), \cdots,\bar{\mathbf{f}}_{K,\lambda}(\mathbf{L}_{k,\lambda}) \}$, $\forall \lambda \in \{t, t-1, \cdots, t-\tau\}$.
Generally, the problem in (\ref{P1}) is challenging since (i) $\bar{\mathbf{\Phi}}_{t}$ and $\mathbf{W}_t$ are sophisticatedly coupled in the objective function and (ii) it is intractable to derive a closed-form expression for the objective function.
In the following, we will propose a predictive beamforming scheme to solve problem (\ref{P1}) sub-optimally.

\begin{figure*}[t]
  \centering
  \includegraphics[width=\linewidth]{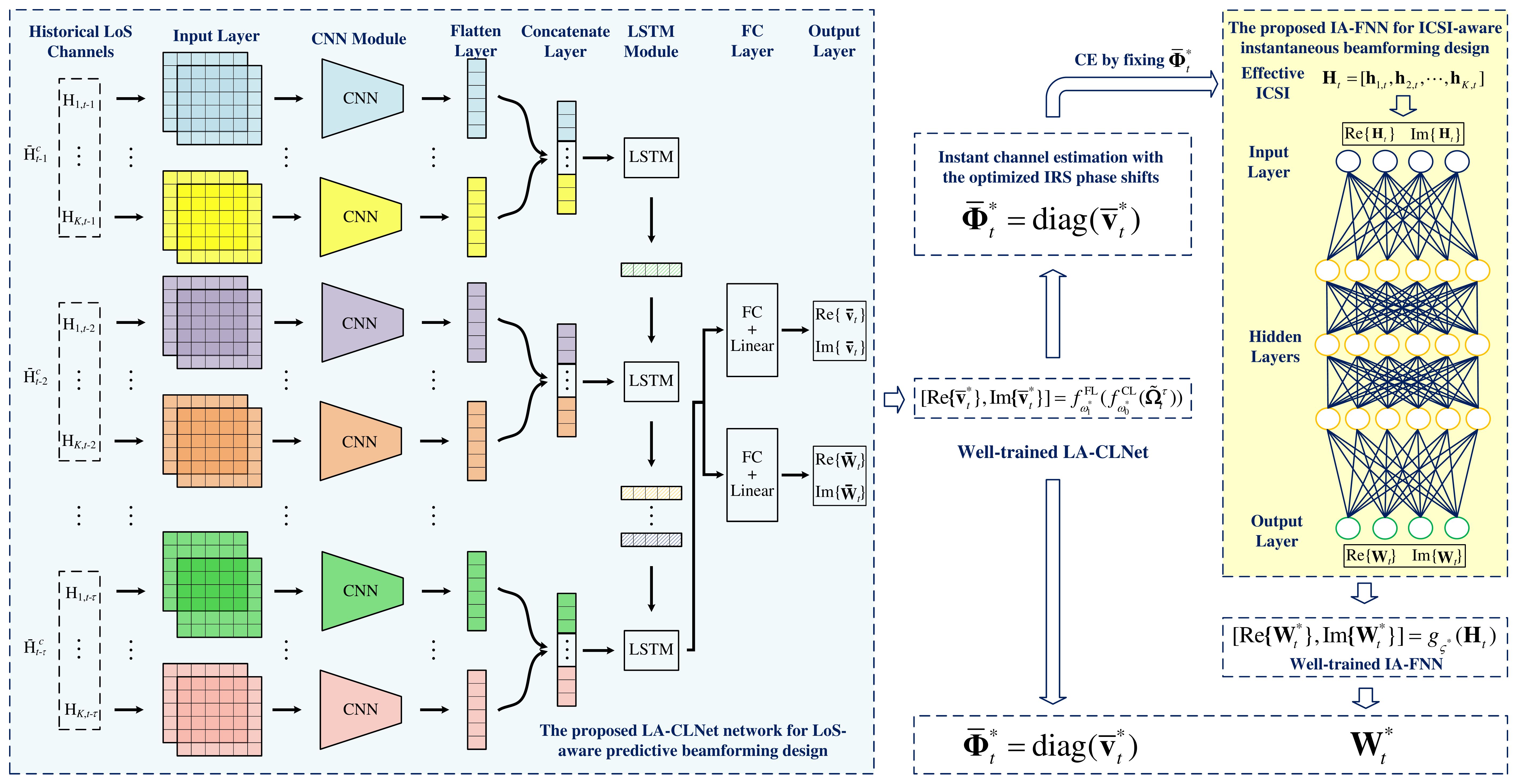}
  \caption{ The proposed LA-CLNet and IA-FNN for predictive beamforming in IRS-MUC systems. }\label{Fig:LA-CLNet}
\end{figure*}

\section{Predictive Beamforming for IRS-MUC Systems}
In this section, we propose an unsupervised DL-based predictive beamforming scheme, where an LA-CLNet is first designed to address the outer maximization in (\ref{P1}) for acquiring the optimized $\bar{\mathbf{\Phi}}_t^*$, and an IA-FNN is then developed to address the inner maximization in (\ref{P1}) to obtain the optimized $\mathbf{W}_t^*$ exploiting the ICSI with $\bar{\mathbf{\Phi}}_t^*$.
By doing this, problem (\ref{P1}) can be sub-optimally addressed in a separated manner, i.e., an LoS-based maximization in Section III-A and an ICSI-based maximization in Section III-B.

\subsection{ LA-CLNet for LoS-aware Predictive Beamforming Design }
Denote by $\bar{\mathbf{H}}_{k,\lambda}^c = \mathrm{diag}(\bar{\mathbf{f}}_{k,\lambda}(\mathbf{L}_{k,t}))\bar{\mathbf{G}} \in \mathbb{C}^{N \times M}$ the cascaded LoS channel of $U_k$ at time slot $\lambda$.
For ease of explosion, we let $\bar{\mathbf{v}}_t = [ e^{j\bar{\varphi}_{1,t}}, e^{j\bar{\varphi}_{2,t}},\cdots, e^{j\bar{\varphi}_{N,t}} ]^T$ denote the diagonal elements of $\bar{\mathbf{\Phi}}_{t}$ and the outer maximization in (\ref{P1}) can be solved sub-optimally through addressing the following problem:
\begin{align}
\max_{\bar{\mathbf{v}}_t,\bar{\mathbf{W}}_t}  &\mathbb{E}_{\bar{\mathbf{H}}_{t}^c|\mathbf{\Omega}_{t}^{\tau}}
\left\{\sum_{k = 1}^K\mathrm{log}_2\left(1 + \frac{|\bar{\mathbf{v}}_t^H\bar{\mathbf{H}}_{k,t}^c\bar{\mathbf{w}}_{k,t}|^2}{\sum_{j \neq k}^K|\bar{\mathbf{v}}_t^H\bar{\mathbf{H}}_{k,t}^c\bar{\mathbf{w}}_{j,t}|^2 + \sigma_k^2} \right)\right\} \label{PA} \\
\mathrm{s.t.}&\sum_{k = 1}^K\|\bar{\mathbf{w}}_{k,t}\|^2\leq P, ~ 0 \leq \bar{\varphi}_{n,t} \leq 2\pi, \forall n \in \mathcal{N}, t \in \mathcal{T} . \notag
\end{align}
Here, $\mathbb{E}_{\bar{\mathbf{H}}_{t}^c|\mathbf{\Omega}_{t}^{\tau}}\{\cdot\}$ is taken over all random realizations of $\bar{\mathbf{H}}_{t}^c$ given the historical cascaded LoS channels  $\mathbf{\Omega}_{t}^{\tau} \triangleq [\bar{\mathbf{H}}^c_{t-1},\cdots,\bar{\mathbf{H}}^c_{t-\tau}] \in \mathbb{C}^{N \times \tau KM}$, where $\bar{\mathbf{H}}^c_{\lambda} \triangleq [\bar{\mathbf{H}}_{1,\lambda}^c, \cdots, \bar{\mathbf{H}}_{K,\lambda}^c] \in \mathbb{C}^{N \times KM}$.


Note that it is generally intractable, if not impossible, to derive the closed-form objective function of (\ref{PA}) since only the historical locations are available. As an alternative, we propose an LA-CLNet to implicitly learn the channel features and directly predict the beamforming matrix.
As shown in Fig. \ref{Fig:LA-CLNet}, the proposed LA-CLNet consists of one input layer, one CNN module, one flatten layer, one concatenate layer, one LSTM module, one fully-connected (FC) layer, and one output layer.
The hyperparameters of LA-CLNet are shown in Table \ref{Tab:Hyperparameters LA-CLNet}.
\begin{table}[t]
\normalsize
\caption{Hyperparameters of the proposed LA-CLNet}\label{Tab:Hyperparameters LA-CLNet}
\centering
\small
\renewcommand{\arraystretch}{1.6}
\begin{tabular}{c c c}
  \hline
   \multicolumn{3}{l}{\textbf{Input}: $\tilde{\mathbf{\Omega}}_{t}^{\tau}$ with the size of $\tau \times K \times N \times M \times 2$}  \\
  \hline
  \hspace{0.1cm} \textbf{Layers/Modules} & \textbf{Parameters} &  \hspace{0.3cm} \textbf{Values}   \\
  \hspace{0.1cm} CNN module & Size of filters & \hspace{0.3cm}  $ 4 \times 3 \times 3 \times 2$   \\
  \hspace{0.1cm} LSTM module & Size of output  & \hspace{0.3cm}  $ 64 \times 1 $   \\
  \hspace{0.1cm} FC layer & Activation function & \hspace{0.3cm} Linear \\
  \hline
   \multicolumn{3}{l}{\textbf{Output}: $[ \mathrm{Re}\{\bar{\mathbf{v}}_t\}, \mathrm{Im}\{\bar{\mathbf{v}}_t\} ]$ and $[ \mathrm{Re}\{\bar{\mathbf{W}}_t\}, \mathrm{Im}\{\bar{\mathbf{W}}_t\} ]$}  \\
  \hline
\end{tabular}
\end{table}
Here, ``Linear'' denotes the linear activation function and the CNN module refers to one convolutional operation with the rectified linear unit (ReLU) activation function and the max pooling operation.
Specifically, the concatenate layer combines the extracted features from different users and send them to the LSTM module for temporal feature exploration.
To exploit the temporal features of $K$ users, we stack the $\tau$ numbers of historical channel matrices of all the users into one matrix with the size of $\tau \times K \times N \times M$ and adopt two neural network channels for the real part and the imaginary part of the complex-valued input, respectively, i.e., the LA-CLNet input can be expressed as $\tilde{\mathbf{\Omega}}_{t}^{\tau} = \mathcal{F}([ \mathrm{Re}\{\mathbf{\Omega}_{t}^{\tau}\}, \mathrm{Im}\{\mathbf{\Omega}_{t}^{\tau}\} ])$,
where $\mathcal{F}(\cdot)\!\!:\mathbb{R}^{N \times 2\tau KM}\mapsto\mathbb{R}^{\tau \times K \times N \times M \times 2}$ is the mapping function.
Thus, the LA-CLNet can be formulated as
\begin{equation}\label{vt}
 f^{\mathrm{FL}}_{\omega_1}(f^{\mathrm{CL}}_{\omega_0}(\tilde{\mathbf{\Omega}}_{t}^{\tau})) = [ \mathrm{Re}\{\bar{\mathbf{v}}_t\}, \mathrm{Im}\{\bar{\mathbf{v}}_t\} ]
\end{equation}
and
\begin{equation}\label{Wt}
 f^{\mathrm{FL}}_{\omega_2}(f^{\mathrm{CL}}_{\omega_0}(\tilde{\mathbf{\Omega}}_{t}^{\tau})) = [ \mathrm{Re}\{\bar{\mathbf{W}}_t\}, \mathrm{Im}\{\bar{\mathbf{W}}_t\} ] ,
\end{equation}
where $f^{\mathrm{CL}}_{\omega_0}(\cdot)$ denotes the common layers (i.e., from the input layer to the LSTM module) with the common parameter $\omega_0$ and $f^{\mathrm{FL}}_{\omega_i}(\cdot)$, $i \in \{1,2\}$, is the FC layer with the parameter $\omega_i$, i.e., $\omega_1$ is for $\bar{\mathbf{v}}_t$ and $\omega_2$ is for $\bar{\mathbf{W}}_t$.

Given an unlabeled training set: $\mathcal{X} = \{ ( \tilde{\mathbf{\Omega}}_k^{\tau(1)},\bar{\mathbf{H}}_{t}^{c (1)} ),  \cdots, ( \tilde{\mathbf{\Omega}}_k^{\tau(N_t)},\bar{\mathbf{H}}_{t}^{c (N_t)} )  \}$,
where $( \tilde{\mathbf{\Omega}}_k^{\tau(i)},\bar{\mathbf{H}}_{t}^{c (i)} )$ is the $i$-th, $i \in \{1,2,\cdots,N_t\}$, training example of $\mathcal{X}$.
Based on the problem formulation in (\ref{PA}), the cost function of LA-CLNet can be expressed as
\begin{equation}\label{J_LA-CLNet}
  J_{\mathrm{LA-CLNet}}(\omega) = -\frac{1}{N_t}\sum_{i=1}^{N_t} \sum_{k = 1}^K\mathrm{log}_2(1 + \bar{\gamma}_{k,t}^{(i)}({\omega}))
\end{equation}
with
$\bar{\gamma}_{k,t}^{(i)}({\omega}) = \frac{|[\bar{\mathbf{v}}_t^{(i)}(\omega)]^H\bar{\mathbf{H}}_{k,t}^{c(i)}\bar{\mathbf{w}}_{k,t}^{(i)}(\omega)|^2}{\sum_{j \in \mathcal{K}\setminus k}|[\bar{\mathbf{v}}_t^{(i)}(\omega)]^H\bar{\mathbf{H}}_{k,t}^{c(i)}\bar{\mathbf{w}}_{j,t}^{(i)}(\omega)|^2 + \sigma_k^2}$,
where $\omega = \{\omega_0,\omega_1,\omega_2\}$ is the network parameters, $\bar{\mathbf{v}}_t^{(i)}(\omega)$ and $\bar{\mathbf{w}}_{k,t}^{(i)}(\omega)$ denote the complex-valued vectors based on the output of LA-CLNet with an input of $\tilde{\mathbf{\Omega}}_k^{\tau(i)}$, respectively.
After offline training via the backpropagation algorithm (BPA), the well-trained LA-CLNet is obtained and we have
\begin{equation}\label{vt*}
 [ \mathrm{Re}\{\bar{\mathbf{v}}_t^*\}, \mathrm{Im}\{\bar{\mathbf{v}}_t^*\} ] = f^{\mathrm{FL}}_{\omega_1^*}(f^{\mathrm{CL}}_{\omega_0^*}(\tilde{\mathbf{\Omega}}_{t}^{\tau}))
\end{equation}
and $[ \mathrm{Re}\{\bar{\mathbf{W}}_t^*\}, \mathrm{Im}\{\bar{\mathbf{W}}_t^*\} ]  = f^{\mathrm{FL}}_{\omega_2^*}(f^{\mathrm{CL}}_{\omega_0^*}(\tilde{\mathbf{\Omega}}_{t}^{\tau}))$,
where $\bar{\mathbf{v}}_t^*$ and $\bar{\mathbf{W}}_t^*$ are the well-trained phase shifts and the beamforming matrix. $\{\omega_0^*, \omega_1^*, \omega_2^*\}$ are the well-trained network parameters. In addition, we omit $\bar{\mathbf{W}}_t^*$ in Fig. \ref{Fig:LA-CLNet} since it is an auxiliary parameter for optimizing $\bar{\mathbf{v}}_t^*$.

\begin{table}[t]
\normalsize
\caption{Hyperparameters of the proposed IA-FNN}\label{Tab:Hyperparameters IA-FNN}
\centering
\small
\renewcommand{\arraystretch}{1.6}
\begin{tabular}{c c c}
  \hline
   \multicolumn{3}{l}{\textbf{Input}: $\mathbf{H}_t$ with the size of $M \times K \times 2$}  \\
  \hline
  \hspace{0.1cm} \textbf{Layers} & \textbf{Parameters} &  \hspace{0.3cm} \textbf{Default Values}   \\
  \hspace{0.1cm} Hidden Layers & Sizes of weights & \hspace{0.3cm}  $32 \times 16 \times 16$   \\
  \hline
   \multicolumn{3}{l}{\textbf{Output}: $[\mathrm{Re}\{{\mathbf{W}}_t\},\mathrm{Im}\{{\mathbf{W}}_t\}]$ }    \\
  \hline
\end{tabular}
\end{table}

\subsection{ IA-FNN for ICSI-aware Instantaneous Beamforming Design }

Given $\bar{\mathbf{\Phi}}_t^* = \mathrm{diag}(\bar{\mathbf{v}}_t^*)$, we only need to obtain the effective ICSI $\mathbf{H}_t = [\mathbf{h}_{1,t},\mathbf{h}_{2,t},\cdots,\mathbf{h}_{K,t}] \in \mathbb{C}^{M \times K}$, where $\mathbf{h}_{k,t}^H \triangleq \mathbf{f}_{k,t}^H\bar{\mathbf{\Phi}}_t^*\mathbf{G}_t \in \mathbb{C}^{1 \times M}$ is an equivalent MISO channel of AP-IRS-$U_k$ link and can be acquired via some existing CE methods \cite{liu2020deepresidual} at Phase I of each time slot\footnotemark\footnotetext{
To focus on the beamforming design, we assume for simplicity that the effective ICSI is perfectly known by the AP as only a MISO channel estimation is required for each user.}.
In this case, we aim to design an ICSI-aware beamforming matrix $\mathbf{W}_t$ to further enhance the sum-rate of the system.
Thus, the inner maximization problem in (\ref{P1}) can be formulated as
\begin{align}
\max_{{\mathbf{W}}_t} \,\,& \sum_{k = 1}^K\mathrm{log}_2\left(1 + \frac{|\mathbf{h}_{k,t}^H\mathbf{w}_{k,t}|^2}{{\sum_{j \neq k}^K}|\mathbf{h}_{k,t}^H\mathbf{w}_{j,t}|^2 + \sigma_k^2} \right) \label{P2} \\
\mathrm{s.t.}\,\,&\sum_{k = 1}^K\|\mathbf{w}_{k,t}\|^2\leq P,  t \in \mathcal{T} \vspace{-0.2cm} . \notag
\end{align}

To handle this problem, we propose an IA-FNN to optimize $\mathbf{W}_t$ exploiting the effective ICSI $\mathbf{H}_t$.
As shown in the right hand side of Fig. \ref{Fig:LA-CLNet}, the proposed IA-FNN consists of an input layer, several hidden layers, and an output layer.
The hyperparameters of IA-FNN are shown in Table \ref{Tab:Hyperparameters IA-FNN}.
Similar to the LA-CLNet, the input of IA-FNN can be expressed as $\mathbf{H}_{t} = \mathcal{G}([ \mathrm{Re}\{\mathbf{H}_{t}\}, \mathrm{Im}\{\mathbf{H}_{t}\} ])$,
where $\mathcal{G}(\cdot)\!\!:\mathbb{R}^{M \times 2K}\mapsto\mathbb{R}^{\tau \times M \times K \times 2}$ is the mapping function.
In this case, the IA-FNN can be formulated as \vspace{-0.1cm}
\begin{equation}\label{}
  g_\varsigma (\mathbf{H}_t) = [\mathrm{Re}\{{\mathbf{W}}_t\},\mathrm{Im}\{{\mathbf{W}}_t\}] ,
\end{equation}
where $g_\varsigma(\cdot)$ is the mathematical function of IA-FNN with parameter $\varsigma$.
Given the effective ICSI $\mathbf{H}_t$ as the training example, the cost function of IA-FNN can be expressed as
\begin{equation}\label{J_IA-FNN}
  J_{\mathrm{IA-FNN}}(\varsigma) = -\sum_{k = 1}^K\mathrm{log}_2 (1 + \tilde{\gamma}_{k,t}({\varsigma})  )
\end{equation}
with
$\tilde{\gamma}_{k,t}({\varsigma}) = \frac{|\mathbf{h}_{k,t}^H\mathbf{w}_{k,t}(\varsigma)|^2}{{\sum_{j \neq k}^K}|\mathbf{h}_{k,t}^H\mathbf{w}_{j,t}(\varsigma)|^2 + \sigma_k^2}$,
where $\mathbf{w}_{k,t}(\varsigma)$ is a complex-valued vector based on the output of IA-FNN with an input of ${\mathbf{H}}_t$.
After online training via the BP algorithm, we can finally obtain the well-trained IA-FNN. Thus, the well-trained beamforming matrix ${\mathbf{W}}_t^*$ can be expressed as
\begin{equation}\label{}
  [\mathrm{Re}\{{\mathbf{W}}_t^*\},\mathrm{Im}\{{\mathbf{W}}_t^*\}] = g_{\varsigma^*} (\mathbf{H}_t),
\end{equation}
where $\varsigma^*$ denotes the well-trained network parameters.

\subsection{ DL-based Predictive Beamforming Algorithm }
Based on Sections III-A and III-B, the proposed DL-based predictive beamforming (DLPB) algorithm is summarized in Algorithm 1, which consists of offline training, online prediction, and online optimization.
Here, $i$ and $j$ are the iteration indices of offline training and online optimization, respectively.
In addition, $I$ and $J$ are the maximum iteration numbers of offline training and online optimization, respectively.

\begin{table}[t]
\small
\centering
\begin{tabular}{l}
\toprule[1.8pt] \vspace{-0.3cm}\\
\hspace{-0.1cm} \textbf{Algorithm 1} { DL-based Predictive Beamforming Algorithm } \vspace{0.1cm} \\
\toprule[1.8pt] \vspace{-0.3cm}\\
\textbf{Initialization:} $i = j = 0$, unlabeled training set $\mathcal{X}$ \\
\textbf{Offline Training:} \\
1:\hspace{0.25cm}\textbf{Input:} Training set $\mathcal{X}$ for LA-CLNet \\
2:\hspace{0.4cm}\textbf{while} $i \leq I $ \textbf{do} \\
3:\hspace{0.8cm}Update $\omega$ by BPA to minimize $J_{\mathrm{LA-CLNet}}(\omega)$ in (\ref{J_LA-CLNet})\\
\hspace{1cm} $i = i + 1$  \\
4:\hspace{0.4cm}\textbf{end while} \\
5:\hspace{0.25cm}\textbf{Output}:  Well-trained $f^{\mathrm{FL}}_{\omega_1^*}(f^{\mathrm{CL}}_{\omega_0^*}(\cdot))$ and $f^{\mathrm{FL}}_{\omega_2^*}(f^{\mathrm{CL}}_{\omega_0^*}(\cdot))$ \\
\textbf{Online Prediction:} (LoS-aware Predictive Beamforming) \\
6:\hspace{0.25cm}\textbf{Input:} Test data $\tilde{\mathbf{\Omega}}_{t}^{\tau} $  \\
7:\hspace{0.4cm}\textbf{do} predictive beamforming via the well-trained LA-CLNet  \\
8:\hspace{0.25cm}\textbf{Output:} $[ \mathrm{Re}\{\bar{\mathbf{v}}_t^*\}, \mathrm{Im}\{\bar{\mathbf{v}}_t^*\} ] = f^{\mathrm{FL}}_{\omega_1^*}(f^{\mathrm{CL}}_{\omega_0^*}(\tilde{\mathbf{\Omega}}_{t}^{\tau}))$ \\
\textbf{Online Optimization:} (ICSI-aware Instantaneous Beamforming) \\
9:\hspace{0.25cm}\textbf{Input:} Test data $\mathbf{H}_{t} $ based on $\bar{\mathbf{v}}_t^* $ for IA-FNN \\
10:\hspace{0.3cm}\textbf{while} $j \leq J $ \textbf{do} \\
11:\hspace{0.7cm}Update $\varsigma$ by BPA to minimize $J_{\mathrm{IA-FNN}}(\varsigma)$ in (\ref{J_IA-FNN})  \\
\hspace{1cm} $j = j + 1$  \\
12:\hspace{0.3cm}\textbf{end while} \\
13:\hspace{0.1cm}\textbf{Output:} $[\mathrm{Re}\{{\mathbf{W}}_t^*\},\mathrm{Im}\{{\mathbf{W}}_t^*\}] = g_{\varsigma^*} (\mathbf{H}_t)$ \\
\bottomrule[1.8pt]
\end{tabular}
\end{table}

\section{ Numerical Results }
In this section, we provide simulation results to verify the effectiveness of the proposed scheme in an IRS-MUC system with $M = 6$, $K = 3$, $ N_y = N_z = 6 $.
The distances between two adjacent antenna/IRS elements are set as $\Delta d_\mathrm{A} = \Delta d_{\mathrm{I}y} = \Delta d_{\mathrm{I}z} = \frac{1}{2}\lambda_c$, respectively.
The locations of AP and IRS are set by $\mathbf{L}_\mathrm{A} = [2,0,20]^T~\mathrm{m}$ and $\mathbf{L}_\mathrm{I} = [0,50,25]^T~\mathrm{m}$, respectively.
The users' initial locations are randomly distributed within a rectangle where the coordinates of top left corner and the bottom right corner are $(3,50,0)$ and $(6,60,0)$, respectively.
The movement function is defined in (\ref{movement_function}) and we set $A_1 = 8~\mathrm{m}/\Delta T$, $A_2 = 10~\mathrm{m}/\Delta T$, $B_1 = -\pi/18$, $B_2 = \pi/18$, $\Delta T = 0.02$ s, and $\sigma_v=0.01$ \cite{9076668}.
In addition, the path losses are modeled as
$\alpha^{\mathrm{AI}} = \beta_0(d^{\mathrm{AI}}/D_0)^{-\eta_{\mathrm{AI}}}$ and $\alpha^{\mathrm{IU}}_{k,t} = \beta_0(d^\mathrm{IU}_{k,t}/D_0)^{-\eta_{k}}$, where $D_0 = 1~\mathrm{m}$ is the reference distance, $\beta_0 = -30~\mathrm{dB}$ is the path loss at $D_0$, $\eta_{\mathrm{AI}} = 2.2$ and $\eta_k = 3$ denote the path loss exponents of AP-IRS and IRS-$U_k$, respectively.
The noise variance are set as $\sigma_k^2 = -96~\mathrm{dBm}$.
For the proposed DLPB algorithm, we set $\tau = 5$ and $N_t = 2,000$.
The hyperparameters of LA-CLNet and IA-FNN are given in Table \ref{Tab:Hyperparameters LA-CLNet} and Table \ref{Tab:Hyperparameters IA-FNN}, respectively.
In addition, three benchmarks have been considered: the (upper bound) genie-aided method, i.e., the DL method in \cite{gao2020unsupervised} with perfect ICSI (denoted by DL-ICSI), the naive DL scheme (denoted by Naive DL), where the LoS channels at time slot $t-1$ are directly adopted for time slot $t$, and the random phase-shift scheme (denoted by Random PS) \cite{guo2020weighted}.
All the simulation results are obtained by averaging over 2,000 Monte Carlo realizations.

\begin{figure}[t]
  \centering
  \includegraphics[width=\linewidth]{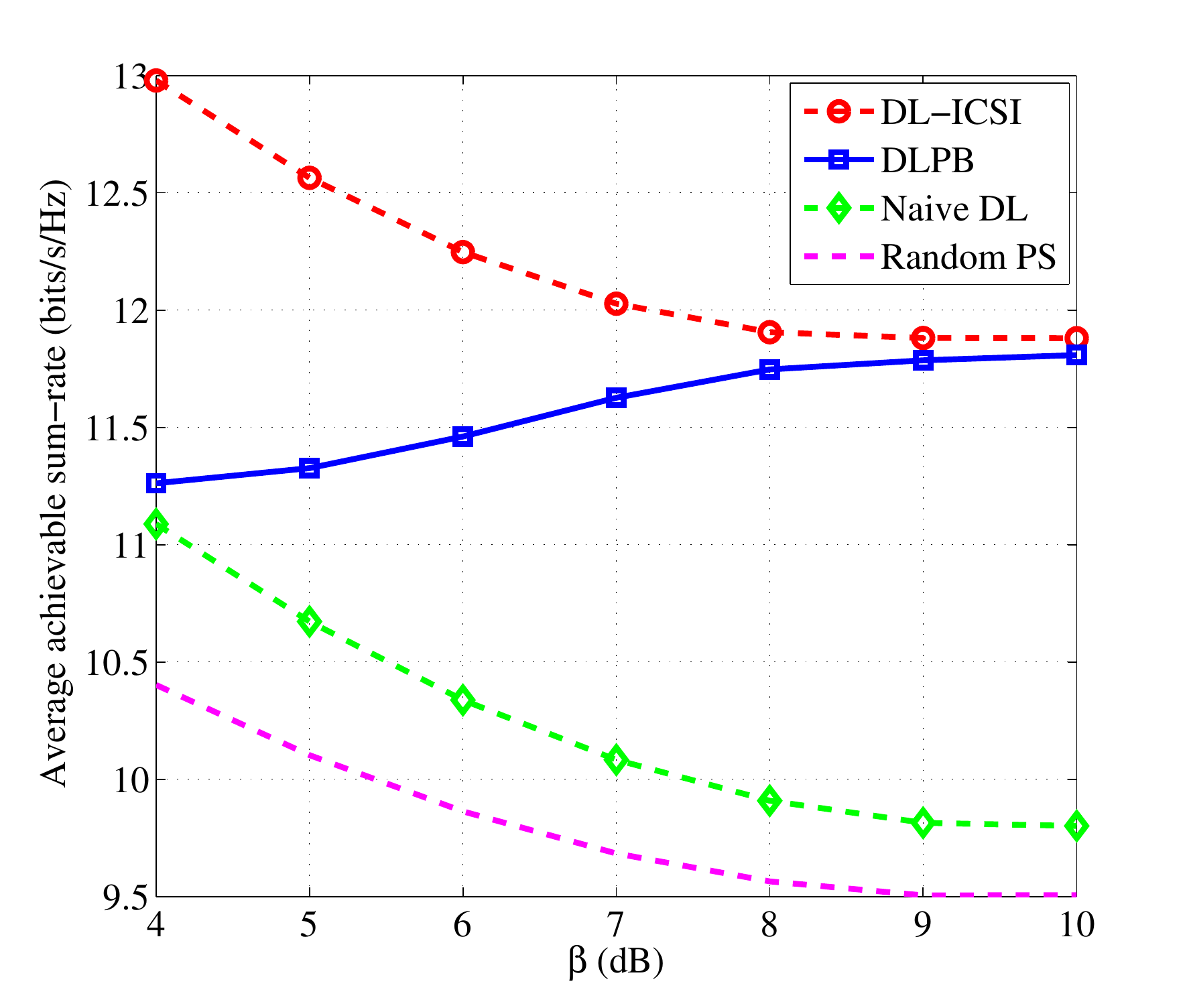}  
  \caption{ The average sum-rate versus $\beta$ with $P = 30~\mathrm{dBm}$. }\label{Fig:sum_rate_beta}
\end{figure}

Fig. \ref{Fig:sum_rate_beta} investigates the average achievable sum-rate versus the Rician factor. For ease of study, we assume  $\beta^{\mathrm{AI}} = \beta^{\mathrm{IU}}_{k,t} = \beta, \forall k,t$.
It can be observed that the average achievable sum-rate of the genie-aided DL-ICSI method decreases with the increasing of $\beta$.
The reason is that when $\beta$ is large, the channels of AP-IRS-user links become more correlated, i.e., the number of degrees-of-freedom of the channels shrinks, thus degrading the data multiplexing capability of the AP for serving multiple users concurrently.
Different from the DL-ICSI method, the average sum-rate of the proposed DLPB method increases with $\beta$ and approaches closely to that of the upper bound when $\beta = 10~\mathrm{dB}$.
This is because when $\beta$ is sufficiently large, the deterministic LoS components of the AP-IRS-user links are dominant, which is beneficial for the beam alignment in our proposed method as the predictive IRS beamforming exploiting the historical information of  LoS channels.
On the other hand, the proposed DLBP method outperforms the naive DL method and the random PS method significantly.
This is expected since the naive method only adopts the outdated ICSI and the random PS method can hardly take advantages of the IRS as the random beamforming at the IRS does not always align with the channels. In contrast, our proposed DLBP method can make the full use of the historical LoS channels and the effective ICSI to further improve the performance.

\begin{figure}[t]
  \centering
  \includegraphics[width=\linewidth]{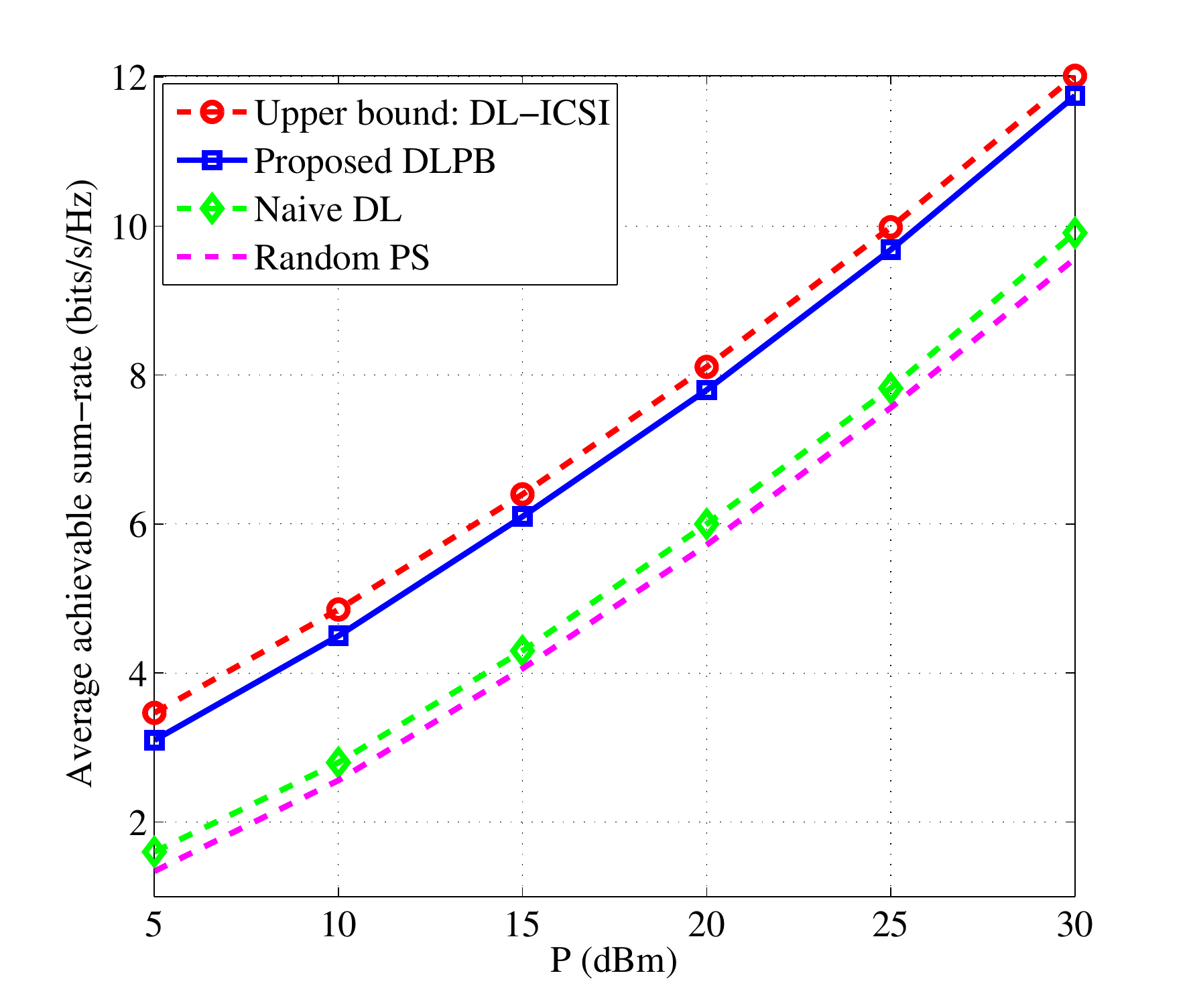}
  \caption{ The average sum-rate versus $P$ with $\beta = 8~\mathrm{dB}$. }\label{Fig:sum_rate_power}
\end{figure}

Besides, the curves of average achievable sum-rate versus the transmit power are presented in Fig. \ref{Fig:sum_rate_power}.
It is shown that the sum-rate of the proposed DLPB method increases with $P$ with the same slope as the DL-ICSI method. Besides, the proposed method can achieve a $5~\mathrm{dB}$ performance gain compared with the naive DL and random PS schemes. The reason is that a predictive IRS phase shift matrix in the DLPB method is much more effective than the non-predictive and random IRS phase shift matrices adopted in the other two methods, respectively.

\section{Conclusion}
This paper proposed a DL-based predictive beamforming scheme to implicitly learn the channel features and directly predict the beamforming vectors for IRS-MUC systems.
In the proposed scheme, a novel LA-CLNet was first developed to efficiently exploit both the spatial and temporal features of the historical LoS channels to predict the IRS phase shifts.
An IA-FNN was then proposed to instantaneously optimize the beamforming matrix at the AP exploiting the effective ICSI with the predictive IRS phase shifts.
Simulation results showed that the proposed scheme can achieve almost the same sum-rate performance as that of the DL method with the full ICSI.

\vspace{0.16cm}

\bibliographystyle{ieeetr}

\setlength{\baselineskip}{10pt}

\bibliography{ReferenceSCI2}

\end{document}